\documentclass[prd,showpacs,showkeys,nofootinbib,floatfix,eqsecnum,fleqn,
                preprint,12pt,tightenlines]{revtex4} %for preprint

\usepackage{amsmath,amssymb,revsymb,graphicx,dcolumn}
\usepackage{comment}
\newcommand  {\version}{v1.00}

\newcommand{\beq}{\begin{equation}}
\newcommand{\eeq}{\end{equation}}
\newcommand{\beqa}{\begin{eqnarray}}
\newcommand{\eeqa}{\end{eqnarray}}
\newcommand{\bsubeqs}{\begin{subequations}}
\newcommand{\esubeqs}{\end{subequations}}

\newcommand{\half}{{\textstyle \frac{1}{2}}}    % 1/2

\begin{document}

\hfill arXiv:1763650 [gr-qc] 08.02.2017 (\version)
                               %%(\version; \today)
\newline\vspace*{5mm}
\title{Four-form field versus fundamental scalar field \vspace*{5mm}}
\author{M.K. Savelainen}
\email{matti.savelainen@aalto.fi}
\affiliation{\mbox{Low Temperature Laboratory, Department of Applied Physics, Aalto University,}\\
PO Box 15100, FI-00076 Aalto, Finland\\
}

\begin{abstract}
\vspace*{2.5mm}\noindent
A modified-gravity theory with a four-form
field strength $F$, a variable gravitational coupling parameter $G(F)$,
and a standard matter action is considered here. Maxwell and Einstein equations are now derived when including to action also derivates of $F$. The energy momentum tensor of the 4-form field contains both the part, which is typical for the fundamental (pseudo)scalar, and the part, which cancels 
the divergent contribution of the zero-point energies of quantum fields to the vacuum energy and thus leads to the natural nullification of the cosmological constant in Minkowski vacuum. 
\end{abstract}

\maketitle

\section{Introduction}\label{sec:introduction}

The physics of the many-body condensed systems provides numerous hints for high energy physics and cosmology. For example, the Gor'kov theory of superconductivity  \cite{Gorkov1958} opened the route to the construction of the relativistic quantum field models,\cite{Nambu1961,VaksLarkin1961,Higgs1964} in which the Higgs bosons are composite objects being analogs of the amplitude modes in superconductors.\cite{Anderson1963}
The connection between the topologically protected Weyl fermions in topological materials --  superfluids, semimetals and superconductors of the Weyl type -- and chiral particles in Standard Model of particle physics also suggests that the Standard Model is an effective theory, where the Weyl fermions, gauge fields and gravity emerge in the vicinity of the topologically protected Weyl points in the spectrum of the quantum vacuum.\cite{FrogNielBook,Horava2005,Volovik2003}

Another connection between the ground state of the many-body condensed matter system and the quantum vacuum is revealed when one  considers  the energy of the quantum vacuum, which contributes to the cosmological constant.
The discrepancy between the observed almost zero value of the vacuum energy and its estimation in terms of the zero point energy of fermionic and bosonic quantum fields provides the cosmological constant problem. Most plausibly the huge discrepancy of about 120 orders is the result of the estimations, which have been based  on low-energy effective field theory. While the condensed matter teaches us that such quantity, as the ground state energy of the quantum systems, can be computed only within a full microscopic quantum theory. Within such theories the condensed matter systems demonstrate, that if the given system is close to equilibrium, the properly determined thermodynamic energy is close to zero. In a full equilibrium the huge contribution of the zero point motion is completely cancelled by the microscopic (correspondingly trans-Planckian) degrees of freedom.  

Unfortunately, to date, we do not have any microscopic theory of the quantum vacuum. However, again the condensed matter demonstrates to us that 
the microscopic degrees of freedom can be also described in terms of their own effective macroscopic variables,  which do not depend mush on the detailed microscopic structure of the system, such as the density of atoms in the many-body quantum systems. The Lorentz invariant analog of such macroscopic approach is represented by the so-called 
$q$-theory. \cite{KlinkhamerVolovik2008,KV2008b,Volovik2013} 
It also provides the general description of the quantum vacuum, and its equations do not depend on the choice of the vacuum variables, and on the microscopic (trans-Planckian) details. As in condensed matter, the sub-Planckian and trans-Planckian contributions to the energy of the fully equilibrium vacuum are naturally  canceled by the thermodynamic argument without fine-tuning, in spite of the huge contribution of the zero point energies.

The particular useful choice for the vacuum variable is the 4-form field, \cite{DuffNieuwenhuizen1980,Aurilia-etal1980,Hawking1984+Duff+Wu, DuncanJensen1989,BoussoPolchinski2000,Aurilia-etal2004}
which satisfies all the requirements needed for the description of the quantum vacuum, especially if instead of the quadratic form in the $F$ field, one use the general function of $F$.
Later it became clear that the $q$-theory must be extended to include  the derivatives of the $q$-field.
In particular this is important for the consideration of the possible origin of dark matter, and for the consideration of the inhomogeneous states of the vacuum, such as interfaces between the vacua.
The latter is important if the Universe is at the coexistence point, where different vacua have the same energy. This is in the origin of the so-called
multiple point principle.\cite{Nielsen2016a,Nielsen2016b,Nielsen2016c,Nielsen1997,Volovik2004}.

This extension was done in Refs. \cite{KV2016-q-DM,KV2016-q-DM2}, where the derivatives of the $q$-field has been added to the action, and it was shown that as a result the oscillations of the $q$-field during cosmological evolution produce the kind of dark matter. In this current paper we, in more detail, derive the Einstein equations that are presented as equations (5) and (6) in the article \cite{KV2016-q-DM}. Also derivation of equation (4) is included. The gravity field is presented here in a more general way like in \cite{KV2008b}.

\section{The action for Gravity with $\boldsymbol{F}$ field with gradients and
         variable gravitational coupling}
\label{sec:Gravity-with-F-field}

The action for the 4-form field interacting with the gravitational field has the following form ($\hbar=c=1$):
\bsubeqs\label{eq:EinsteinF-all}
\beqa
S&= -
\int_{\mathbb{R}^4} \,d^4x\, \sqrt{|g|}\,\left(\frac{R}{16\pi G(F)} +
\epsilon(F)
\right.
\left.
+\frac{1}{8}\,K(F)\,\nabla^{\alpha} F^2\,\nabla_\alpha F^2
+\mathcal{L}^\text{\,SM}\right)\,,
\label{eq:actionF}\\[2mm]
&F_{\kappa\lambda\mu\nu}\equiv
\nabla^{\phantom{(a)}}_{[\kappa}\!\!A^{\phantom{(a)}}_{\lambda\mu\nu]}\,,
\label{eq:Fdefinition}\,\quad
F^2\equiv
-\,\frac{1}{4!}\,F_{\kappa\lambda\mu\nu}F^{\kappa\lambda\mu\nu}\\[2mm]
&F_{\kappa\lambda\mu\nu}=F\sqrt{|g|} \,e_{\kappa\lambda\mu\nu}\,,\quad
F^{\kappa\lambda\mu\nu}=F \,e^{\kappa\lambda\mu\nu}/\sqrt{|g|}\,,
\label{eq:Fdefinition2}
\eeqa
\esubeqs
 $\nabla_\mu$ denotes a covariant derivative
and a square bracket around spacetime indices complete anti-symmetrization. $\nabla^{\alpha} F^2\,\,\nabla_\alpha F^2$ is $g^{\alpha\,\beta}\,\nabla_\beta F^2\,\,\nabla_\alpha F^2$. $K(F)$ is some factor depending on $F$ only (here not on its derivatives). $\mathcal{L}^\text{\,SM}$ is the Lagrange density of the fields
of the standard model (SM) of elementary particle physics.
Throughout, we use natural units with $c=\hbar=1$
and take the metric signature $(-+++)$.

Variation over $A_{\lambda\mu\nu}$ gives the Maxwell equations, see Appendix:
\begin{equation}\label{maxwell}\begin{split}
\nabla_\kappa\,\left(\frac{R}{16\pi}\,\frac{dG^{-1}(F)}{dF} +
\,\frac{d\epsilon(F)}{dF}\,+\,\frac{1}{8}\frac{dK(F)}{dF}\partial^\alpha F^2 \partial_\alpha F^2\,-\,\frac{1}{2}\,F\,\nabla^\alpha\left( K(F)\,\partial_\alpha F^2\right)\right)\,=\,0 \,.
\end{split}\end{equation}
From Maxwell equation  \eqref{maxwell} we get 
\begin{equation}\label{maxwell1112}
\begin{split}
\frac{R}{16\pi}\,\frac{dG^{-1}(F)}{dF} +
\,\frac{d\epsilon(F)}{dF}\,+\,\frac{1}{8}\frac{dK(F)}{dF}\partial^\alpha F^2 \partial_\alpha F^2\,-\,\frac{1}{2}\,F\,\nabla^\alpha\left( K(F)\,\partial_\alpha F^2\right)\,=\mu, 
\end{split}\end{equation}
where $\mu$ is the integration constant. It plays the role of chemical potential, which is thermodynamically conjugate to $F$.
It is convenient to set
\begin{equation}\label{e:111}\begin{split}
C(F)\,=\,F^2\,K(F) \,,
\end{split}\end{equation}
which gives for Maxwell equations:
\begin{equation}\label{maxwell11112}\begin{split}
\frac{R}{16\pi}\,\frac{dG^{-1}(F)}{dF} +
\,\frac{d\epsilon(F)}{dF}\,-\, \frac{1}{2}\frac{dC(F)}{dF}\,\partial^{\alpha} F \partial_{\alpha} F\,-\, C(F) \,\Box F\,\,=\mu. 
\end{split}\end{equation}

Variation over the metric $g^{\mu\nu}$ gives the generalized Einstein equations, see Appendix:
\begin{equation}\label{einstein111}\begin{split}
& \frac{1}{8\pi G(F)}\Big( R_{\mu\nu}-\half\,R\,g_{\mu\nu} \Big)\,+\,\frac{1}{16\pi}\,F\,\frac{dG^{-1}}{dF}\,R\,g_{\mu\nu}\,+ \\[2mm]
& +\,\frac{1}{8\pi}
\Big( \nabla_\mu\nabla_\nu\, G^{-1}(F) - g_{\mu\nu}\, \Box\, G^{-1}(F)\Big)\,-\left(\epsilon (F)-F\,\frac{d\epsilon (F)}{dF}\,\right)g_{\mu\nu}\,+ \\[2mm]
&-\,\frac{1}{2}\,C(F)\,g_{\mu\nu}\,
\nabla_{\alpha} F \,\nabla^{\alpha} F \,-\,\frac{1}{2}\,F\,\frac{dC(F)}{dF}\,\nabla_{\alpha} F \,\nabla^{\alpha} F\,g_{\mu\nu}\,+ \\[2mm]
&+ \,\,C(F)
\nabla_{\mu} F \,\nabla_{\nu} F\,-\,C(F)\,\Box F\,g_{\mu\nu}
+  T_{\mu\nu}^{({\rm SM})}
\,=0\,,
\end{split}\end{equation}
which can be simplified using Eq.(\ref{maxwell11112}):
\begin{equation}
\label{einstein1111}
\begin{split}
& \frac{1}{8\pi G(F)}\Big( R_{\mu\nu}-\half\,R\,g_{\mu\nu} \Big)\,+\,\frac{1}{8\pi}
\Big( \nabla_\mu\nabla_\nu\, G^{-1}(F) - g_{\mu\nu}\, \Box\, G^{-1}(F)\Big)\,+ \\[2mm]
& -\left(\epsilon (F)-\,\mu\,F\,\right)g_{\mu\nu}\,-\,\frac{1}{2}\,C(F)\,g_{\mu\nu}\,
\nabla_{\alpha} F \,\nabla^{\alpha} F \,+\,C(F)
\nabla_{\mu} F \,\nabla_{\nu} F\,
+  T_{\mu\nu}^{({\rm SM})}
=0\,.
\end{split}
\end{equation}
For the constant gravitational coupling $G(F)$ these equations are reduced to the corresponding equation in the article \cite{KV2016-q-DM},

The Einstein equation (\ref{einstein1111}) shows that the contribution of the 4-form field to the gravitating energy-momentum tensor is given by
\begin{eqnarray}
 \label{eq:EM-q}
  T_{\alpha\beta}^{\,(F)}
 =
\left(  C(F) \nabla_\alpha \, F \, \nabla_\beta \, F - 
\frac12 g_{\alpha\beta} C(F) \, \nabla_\mu\, F \, \nabla^\mu F\right) 
-\, g_{\alpha\beta} \left( \epsilon(F) - \mu \, F\right)  \,.
\end{eqnarray} 
The first term on the RHS of Eq.(\ref{eq:EM-q}) corresponds to the energy-momentum tensor  of the conventional scalar field. However, the second term on the RHS of Eq.(\ref{eq:EM-q})
demonstrates the consequence of the fact that $F$ is not a fundamental (pseudo)scalar but a composite
object made of the gauge field $A_{\kappa\lambda\mu}$
and the metric $g_{\mu\nu}$.

\section{Conclusion}\label{sec:conclusion}

There are two faces of the 4-form field, as follows from Eq.(\ref{eq:EM-q}): it has the signature of the (pseudo)scalar and the signature of the conserved quantity, which characterizes the deep quantum vacuum. 
Due to the latter, the contribution to the vacuum energy from the $F$-field is $\epsilon(F) - \mu F$, instead of  the conventional term  $\epsilon(F)$ in the theory of the fundamental scalar field. This difference allows us to avoid the fine-tuning problem \cite{Weinberg1988} in estimation of the contribution of the vacuum energy to the cosmological constant in Minkowski vacuum. While in the scalar field approach the nullification of the diverging contribution of zero-point-energies to the vacuum energy density $\epsilon(F)$ looks artificial, in the $F$-theory the term $\epsilon(F)$ is automatically 
cancelled by the counter term $-\mu F$. This cancellation
is required by the Gibbs--Duhem identity, which is applicable to any equilibrium ground state, including the one of the physical vacuum. As a result the diverging contribution of zero-point-energies of quantum fields is fully compensated by microscopic degrees of freedom, which are effectively described by the vacuum field $F$. The proper vacuum energy density entering the Einstein gravitational equation (\ref{einstein1111}) as cosmological constant is zero in a full equilibrium at zero temperature, $\Lambda=\epsilon(F)-\mu F=0$.

The same nullification happens if in addition to the $F$-field there is the vacuum contribution of the matter fields \cite{KlinkhamerVolovik2008}. For example, if there is the  fundamental scalar field $\Phi$ in the matter sector with the vacuum energy density $\epsilon_{\rm matter}(\Phi)$, the total vacuum energy in equilibrium will be also cancelled, 
$\Lambda=\epsilon(F) + \epsilon_{\rm matter}(\Phi)-\mu F=0$. The compensation of the energy density comes again from the microscopic degrees of freedom. The chemical potential $\mu$ of the vacuum field is self-tuned to the matter fields.

It is important that the form of the Einstein equation in terms of the $F$-field  (\ref{einstein1111})
and the form of the equation \eqref{maxwell11112} for the $F$-field are general, and do not depend much on the origin of the vacuum field. The only specific property of the $F$-field, which leads to the counter term $-\mu F$ in the cosmological constant, is the appearance of chemical potential $\mu$ for $F$-field in \eqref{maxwell11112}. The existence of the chemical potential in the system is the typical consequence of the conservation law. Thus the form of the equations is the consequence of the conservation law, and does not depend on the particular choice of the vacuum field. The $F$-field is the particular representation of the  vacuum field, which obey the proper conservation law.

The obtained equations are applicable for different problems such as: (i) relaxation of the vacuum energy in the expanding Universe; (ii) the internal structure of the black hole including the structure of the singularity;
(iii) investigation of topological and non topological objects; etc.

\section{ACKNOWLEDGMENTS}

I would like to express my deep gratitude to Professor G.V. Volovik for his guidance. This work has been supported by the Academy of Finland  
(Project No. 284594). 

\section*{Appendix}
\label{app}

We divide action into three parts:
\bsubeqs\label{eq:ActionAll}
\beqa
S&=& S_0\,+\,S_1\,+\,S_2\,,
\label{eq:actionOld}\\[2mm]
S_0&=& -
\int_{\mathbb{R}^4} \,d^4x\, \sqrt{|g|}\,\left(\frac{R}{16\pi G(F)} +
\epsilon(F)\right)\!,
\label{eq:actionDeriv}\\[2mm]
S_1&=& -
\int_{\mathbb{R}^4} \,d^4x\, \sqrt{|g|}\,\left(
\frac{1}{8}\,K(F)\,\nabla^{\alpha} F^2\,\nabla_\alpha F^2\right)\!,
\label{eq:actionmatter}\\[2mm]
S_2&=& -
\int_{\mathbb{R}^4} \,d^4x\, \sqrt{|g|}\,\,\mathcal{L}^\text{\,SM}\!.
\eeqa
\esubeqs
Next we variate $S_1$ with respect to the three-form gauge field $A$. Lets first open $S_1$.
\beqa\label{eq:ActionOpen1}\begin{split}
S_1=&\,-
\int_{\mathbb{R}^4} \,d^4x\, \sqrt{|g|}\,
\frac{1}{8}\,K(F)\,
g^{\alpha\beta}\,\times \\[2mm]
&\times\,\nabla_{\beta}\!\left(\!\nabla^{\phantom{(a)}}_{[\kappa}\!\!A^{\phantom{(a)}}_{\lambda\mu\nu]}\,
\,\nabla\,^{[\kappa}\!\!\,A^{\lambda\mu\nu]}\!\right)
\nabla_{\alpha}\!\left(\nabla^{\phantom{(a)}}_{[\kappa}\!\!A^{\phantom{(a)}}_{\lambda\mu\nu]}\,
\!\nabla\,^{[\kappa}\!\!\,A^{\lambda\mu\nu]}\!\right)\frac{1}{4!^{2}}\,.
\end{split}\eeqa
$\underline{\lambda}$, $\underline{\mu}$ and $\underline{\nu}$ are from now on specific index values and usual sum convention not used for them. This underlined set of indices defines the special direction, to which we make the variation $\delta A^{\underline{\lambda\mu\nu}}$. Note that now all other $\delta A^{\lambda\mu\nu}$ and  $\nabla \delta A^{\lambda\mu\nu}$ are zero if $\lambda \ne \underline\lambda$ or $\mu \ne \underline\mu$ or $\nu \ne \underline\nu$. This convention is taken just to make summing in the following derivation easier.

\begin{equation}\label{e:barwq}\begin{split}
S_1\, {}& +\,\delta\, S_1\,=-\int_{\mathbb{R}^4} \,d^4x\, \sqrt{|g|}\,\frac{1}{8}\,K(F+\delta\,F)\cdot\frac{1}{4!^{2}}\times\\[2mm]
&\,\times\,\,g^{\alpha\beta}\nabla_{\beta}\!\left(\!\nabla^{\phantom{(a)}}_{[\kappa}\!\!A^{\phantom{(a)}}_{\lambda\mu\nu}\,+\,\delta\, A^{\phantom{(a)}}_{\underline{\lambda\mu\nu}\,]}\, 
\,\nabla\,^{[\kappa}\!\!\,A^{\lambda\mu\nu}\,+ \delta\, A^{\underline{\lambda\mu\nu}\,]}\!\right)\,\times   \\[2mm]
&  \phantom{aaa}\times\,\,\nabla_{\alpha}\!\left(\nabla^{\phantom{(a)}}_{[\acute\kappa}\!\!A^{\phantom{(a)}}_{\acute\lambda\acute\mu\acute\nu}\,
  + \delta\, A^{\phantom{(a)}}_{\underline{\lambda\mu\nu}\,]}\,\!\nabla\,^{[\acute\kappa}\!\!\,A^{\acute\lambda\acute\mu\acute\nu}+ \delta\, A^{\underline{\lambda\mu\nu}\,]}\!\right) \\[2mm]
&= -\int_{\mathbb{R}^4} d^4x\sqrt{|g|}\frac{1}{8}K(F+\delta\,F)\cdot\frac{1}{4!^{2}}\times \\[2mm]
\times\, &g^{\alpha\beta}\nabla_{\beta}\!\left[\!\nabla^{\phantom{(a)}}_{[\kappa}\!\!A^{\phantom{(a)}}_{\lambda\mu\nu]} 
\nabla^{[\kappa}\!\!A^{\lambda\mu\nu]}+\!\nabla^{\phantom{(a)}}_{[\kappa}\!\!A^{\phantom{(a)}}_{\lambda\mu\nu]}
\nabla^{[\kappa}\delta\,A^{\underline{\lambda\mu\nu}\,]}+\!\nabla^{\phantom{(a)}}_{[\kappa}\!\!\,\delta\, A^{\phantom{(a)}}_{\underline{\lambda\mu\nu}]}\, 
\,\nabla\,^{[\kappa}\!\!\,A^{\lambda\mu\nu]}\,+\,\text{2nd ord}\right]\,\times\\[2mm]
&\times\nabla_{\alpha}\!\left[\!\nabla^{\phantom{(a)}}_{[\acute\kappa}\!\!A^{\phantom{(a)}}_{\acute\lambda\acute\mu\acute\nu]}
\nabla^{[\acute\kappa}\!\!\,A^{\acute\lambda\acute\mu\acute\nu]}+\!\nabla^{\phantom{(a)}}_{[\acute\kappa}\!\!A^{\phantom{(a)}}_{\acute\lambda\acute\mu\acute\nu]}\nabla^{[\acute\kappa}\,\delta\, A^{\underline{\lambda\mu\nu}\,]}+\!\nabla^{\phantom{(a)}}_{[\acute\kappa}\!\!\,\delta A^{\phantom{(a)}}_{\underline{\lambda\mu\nu}]} 
\nabla\,^{[\acute\kappa}\!\!\,A^{\acute\lambda\acute\mu\acute\nu]}+\text{2nd ord}\right]\\[2mm]
\end{split}\end{equation}
\begin{equation}\label{e:barwq1}\begin{split}
S_1\, {}& +\,\delta\, S_1\,
\approx -\int_{\mathbb{R}^4} \,d^4x\, \sqrt{|g|}\,\frac{1}{8}\,\left(K(F)+\frac{dK(F)}{dF} \,\delta F\,\right)\,\times \\[2mm]
&\phantom{aaa}\times\,\,g^{\alpha\beta}\left[\nabla_{\beta}F^2\,+\,2\,\nabla_{\beta}\!\left(\frac{F_{\kappa\lambda\mu\nu}\!\nabla^{\kappa}\delta\, A^{\underline{\lambda\mu\nu}\,}}{-4!}\cdot 24\!\right)\right]\left[\nabla_{\alpha}F^2\,+\,2\,\nabla_{\alpha}\!\left(\frac{F_{\acute\kappa\acute\lambda\acute\mu\acute\nu}\!\nabla^{\acute\kappa}\delta\, A^{\underline{\lambda\mu\nu}\,}}{-4!}\cdot 24\,\!\right)\right] \\[2mm]
&= S_1\,-\int_{\mathbb{R}^4} \,d^4x\, \sqrt{|g|}\,\frac{1}{8}\,\frac{dK(F)}{dF} \,\delta F\,\nabla^{\alpha}F^2\,\nabla_{\alpha}F^2\,+\\[2mm]
  &\phantom{aaa}\,-\int_{\mathbb{R}^4} \,d^4x\, \sqrt{|g|}\,\frac{1}{8}\,K(F)\,g^{\alpha\beta}\left[\,2\,\nabla_{\beta}\!\left(-F_{\kappa\lambda\mu\nu}\!\nabla^{\kappa}\delta\, A^{\underline{\lambda\mu\nu}\,}\!\right)\nabla_{\alpha}F^2\,\right] \,+\,\\[2mm] 
  &\phantom{aaa}\,-\int_{\mathbb{R}^4} \,d^4x\, \sqrt{|g|}\,\frac{1}{8}\,K(F)\,g^{\alpha\beta}\left[\,2\,\nabla_{\alpha}\!\left(-F_{\acute\kappa\acute\lambda\acute\mu\acute\nu}\!\nabla^{\acute\kappa}\delta\, A^{\underline{\lambda\mu\nu}\,}\!\right)\nabla_{\beta}F^2\,\right] +\,\text{2nd ord .} \\[2mm]
\end{split}\end{equation}
For $\delta\,F$ we get
\begin{equation}\label{e:barwq}\begin{split}
\delta\,F\,&=
\,-\,\frac{1}{F}\,\,F_{\kappa\underline{\lambda\mu\nu}}\!\nabla^{\kappa}\delta\, A^{\underline{\lambda\mu\nu}\,}\,.
\end{split}\end{equation}
This gives then
\begin{equation}\label{e:barwq}\begin{split}
\delta\,S_1\,&=\,-\int_{\mathbb{R}^4} \,d^4x\, \sqrt{|g|}\,\frac{1}{8}\,\left[\frac{dK(F)}{dF} \,\,\frac{1}{-F}\,\,F_{\kappa\underline{\lambda\mu\nu}}\!\nabla^{\kappa}\delta\, A^{\underline{\lambda\mu\nu}\,}\,\nabla^{\alpha} F^2\,\nabla_\alpha F^2 \right. \,+ \\[2mm] 
&\left.\,+\,4\,K(F)\,\nabla^{\alpha}\left(-\,F_{\kappa\underline{\lambda\mu\nu}}\nabla^{\kappa}\delta\, A^{\underline{\lambda\mu\nu}\,}\right)\,\nabla_{\alpha}F^{2}\,\right] \\[2mm]
&=\,-\int_{\mathbb{R}^4} \,d^4x\, \sqrt{|g|}\,\frac{1}{8}\,\left[\frac{dK(F)}{dF} \,\,\frac{1}{-F}\,\,F_{\kappa\underline{\lambda\mu\nu}}\!\nabla^{\kappa}\delta\, A^{\underline{\lambda\mu\nu}\,}\,\nabla^{\alpha} F^2\,\nabla_\alpha F^2\,\right. \,+ \\[2mm]
&\left.+\,4\,K(F)\,\partial^{\alpha}\left(-\,F_{\kappa\underline{\lambda\mu\nu}}\nabla^{\kappa}\delta\, A^{\underline{\lambda\mu\nu}\,}\right)\,\nabla_{\alpha}F^{2}\,\right]\\[2mm]
&=\, -\int_{\mathbb{R}^4} \,d^4x\, \sqrt{|g|}\,\frac{1}{8}\,\left[\,\frac{dK(F)}{dF} \,\,\frac{1}{-F}\,\,F_{\kappa\underline{\lambda\mu\nu}}\!\nabla^{\kappa}\delta\, A^{\underline{\lambda\mu\nu}\,}\,\nabla^{\alpha} F^2\,\nabla_\alpha F^2\right]\,+ \delta\,S_{12}
\end{split}\end{equation}

\begin{equation}\label{e:barwq}\begin{split}
\delta\,S_{12}\,=-
\int_{\mathbb{R}^4} \,d^4x\, \sqrt{|g|}\,
\frac{4}{8}\,K(F)\,
g^{\alpha\beta}\left[-\,\partial_{\beta}\!\left(\!\nabla^{\phantom{(a)}}_{\kappa}\!\!\delta\,A^{\phantom{(a)}}_{\underline{\lambda\mu\nu}}\!\right)\,
F\,^{\kappa\underline{\lambda\mu\nu}}\,-\,\!\nabla^{\phantom{(a)}}_{\kappa}\!\!\delta\,A^{\phantom{(a)}}_{\underline{\lambda\mu\nu}}\,
\,\partial_{\beta}F\,^{\kappa\underline{\lambda\mu\nu}}\!\,\right]\,\partial_\alpha\,F^2\,.
\end{split}\end{equation}
The first term of $\delta\,S_{12}$ must be studied in detail. Definition of covariate derivate gives
\begin{equation}\label{e:barwq}\begin{split}
\nabla_\alpha\left(\nabla_\kappa\,\delta\,A_{\underline{\lambda\mu\nu}}\right)\,&=  \,\partial_\alpha\left(\nabla_\kappa\,\delta\,A_{\underline{\lambda\mu\mu}}\,\right)\,- \Gamma^{\acute{\alpha}}_{\alpha\kappa}\,\nabla_{\acute{\alpha}}\,\delta\,A_{\underline{\lambda\mu\nu}} \,-\,\Gamma^{\acute{\alpha}}_{\alpha\underline\lambda}\,\nabla_\kappa\,\delta\,A_{{\acute{\alpha}\underline{\mu\nu}}} \,+\\[2mm]
&-\,\Gamma^{\acute{\alpha}}_{\alpha\underline\mu}\,\nabla_\kappa\,\delta\,A_{\underline{\lambda}\acute\alpha\underline{\nu}}\,-\,\Gamma^{\acute{\alpha}}_{\alpha\underline\nu}\,\nabla_\kappa\,\delta\,A_{\underline{\lambda\mu}\acute\alpha}\,. 
\end{split}\end{equation}
\\
As we assumed that  $\nabla_\kappa\,\delta\,A_{\underline{\lambda\mu\nu}}\,\neq\,0$ only when $\kappa\,\neq\,\underline\lambda\,$,$\,\underline\mu$ and $\underline\nu$\,,
so we have

\begin{equation}\label{e:barwq}\begin{split}
{\nabla_\alpha\left(\nabla_\kappa\,
\delta\,A_{\underline{\lambda\mu\nu}}\right)
\,=\,\partial_\alpha\left(\nabla_\kappa \delta\,A_{\underline{\lambda\mu\mu}}\right)
\,-\,\left(\Gamma^{\underline\kappa}_{\alpha\underline\kappa}\,
\,+\,\Gamma^{\underline\lambda}_{\alpha\underline\lambda}\,+\,\Gamma^{\underline\mu}_{\alpha\underline\mu}\,
\,+\,\Gamma^{\underline\nu}_{\alpha\underline\nu}\right)\,\nabla_\kappa\,\delta\,A_{\underline{\lambda\mu\nu}}
}\,.
\end{split}\end{equation}
Also we have
\\
\begin{equation}\label{e:barwq}\begin{split}
\nabla_\alpha\,F^{\kappa\underline{\lambda\mu\nu}}\,&=\,\partial_\alpha\,F^{\kappa\underline{\lambda\mu\nu}}\,+\,\Gamma^{\kappa}_{\alpha\acute\alpha}\,F^{\acute\alpha\underline{\lambda\mu\nu}}
\,+\,\Gamma^{\underline\lambda}_{\alpha\acute\alpha}\,F^{\kappa\acute\alpha\underline{\mu\nu}}
\,+\,\Gamma^{\underline\mu}_{\alpha\acute\alpha}\,F^{\kappa\underline{\lambda}\acute\alpha\underline{\nu}}
\,+\,\Gamma^{\underline\nu}_{\alpha\acute\alpha}\,F^{\kappa\underline{\lambda\mu}\acute\alpha}\\
&=\,\partial_\alpha\,F^{\kappa\underline{\lambda\mu\nu}}
\,+\left(\Gamma^{\underline\kappa}_{\alpha\underline\kappa}\,
\,+\,\Gamma^{\underline\lambda}_{\alpha\underline\lambda}\,+\,\Gamma^{\underline\mu}_{\alpha\underline\mu}\,
\,+\,\Gamma^{\underline\nu}_{\alpha\underline\nu}\right)\,F^{\kappa\underline{\lambda\mu\nu}}\,.
\end{split}\end{equation}
\\
If we now contract this with $F_{\kappa\underline{\lambda\mu\nu}}$, solve the four-$\Gamma$-factor term and substitute it to the equation for to the $\nabla_\alpha\left(\nabla_\kappa\,\delta\,A_{\underline{\lambda\mu\nu}}\right)\,$ we get finally

\begin{equation}\label{e:barwq11}\begin{split}
\,\partial_\alpha\left(\nabla_\kappa \delta\,A_{\underline{\lambda\mu\mu}}\right)\,F^{\kappa\underline{\lambda\mu\nu}}=&\,\,
\nabla_\alpha\left(\nabla_\kappa\,\delta\,A_{\underline{\lambda\mu\nu}}\right)\,F^{\kappa\underline{\lambda\mu\nu}}\,+\,\nabla_\alpha\,F^{\kappa\underline{\lambda\mu\nu}}\,\nabla_\kappa\,\delta\,A_{\underline{\lambda\mu\nu}}\,+ \\[2mm]
&-\,\partial_\alpha\,F^{\kappa\underline{\lambda\mu\nu}}\,\nabla_\kappa\,\delta\,A_{\underline{\lambda\mu\nu}}
\,.
\end{split}\end{equation}
For $\delta\,S_{12} $ we then get
\begin{equation}\label{e:barwq}\begin{split}
-\delta\,S_{12}\,&=-\int_{\mathbb{R}^4} \,d^4x\, \sqrt{|g|}\,\frac{4}{8}\,K(F)\,
g^{\alpha\beta}\times \\[2mm]
\times&\left[\nabla_{\beta}\!\left(\!\nabla^{\phantom{(a)}}_{\kappa}\!\!\delta\,A^{\phantom{(a)}}_{\underline{\lambda\mu\nu}}\!\right)\,
F\,^{\kappa\underline{\lambda\mu\nu}}\,+\,
\nabla_\beta\,F^{\kappa\underline{\lambda\mu\nu}}\,
\nabla_\kappa\,\delta\,A_{\underline{\lambda\mu\nu}}\,
-\,\partial_{\beta\,}
F^{\kappa\underline{\lambda\mu\nu}}\,
\nabla_\kappa\,\delta\,A_{\underline{\lambda\mu\nu}}\,+\right. \\[2mm]
&\left. +\,\nabla^{\phantom{(a)}}_{\kappa}\!\!\delta\,A^{\phantom{(a)}}_{\underline{\lambda\mu\nu}}\,
\,\partial_{\beta}F\,^{\kappa\underline{\lambda\mu\nu}}\!\,\right]\,\partial_\alpha\,F^2\,.
\end{split}\end{equation}
Due to gauss at the far boundary where $\delta\,A$ can be set to zero we get 
\begin{equation}\label{e:barwq}\begin{split}
-\delta\,S_{12}\,=-\int_{\mathbb{R}^4} \,d^4x\, \sqrt{|g|}\,\frac{4}{8}&\left[\,-\,\nabla^{\phantom{(a)}}_{\kappa}\!\!\delta\,A^{\phantom{(a)}}_{\underline{\lambda\mu\nu}}\,
\nabla^\alpha\left(F\,^{\kappa\underline{\lambda\mu\nu}}\,\partial_\alpha{F^2}\,K(F)\right)\,\right.+ \\[2mm]
&\left.+\,\nabla^{\alpha}\,F^{\kappa\underline{\lambda\mu\nu}}\,\nabla_\kappa\,\delta\,A_{\underline{\lambda\mu\nu}}\,\partial_{\alpha}\,F^2\,K(F)\right] \, \\[2mm]
=\, +\int_{\mathbb{R}^4} \,d^4x\, \sqrt{|g|}\,
\frac{4}{8}&\,F^{\kappa\underline{\lambda\mu\nu}}\,\nabla^\alpha\left(\partial_\alpha\,F^2\,K(F)\right)\nabla_\kappa\,\delta\,A_{\underline{\lambda\mu\nu}}\,. 
\end{split}\end{equation}
Now we study $S_0$ as given in equation \eqref{eq:actionDeriv} and variate it with respect to $\delta\,A_{\underline{\lambda\mu\nu}}\,$.
\begin{equation}\label{e:barwq}\begin{split}
&\delta\,S_0= -\int_{\mathbb{R}^4} \,d^4x\, \sqrt{|g|}\,\,\frac{\partial}{\partial A_{\underline{\lambda\mu\nu}}}\left(\frac{R}{16\pi G(F)} +
\epsilon(F)\right)\,\delta\,A_{\underline{\lambda\mu\nu}}\, \\[2mm]
&=-
\int_{\mathbb{R}^4} \,d^4x\, \sqrt{|g|}\,\,\frac{\partial F}{\partial A_{\underline{\lambda\mu\nu}}}\left(\frac{R}{16\pi}\,\frac{dG^{-1}(F)}{dF} +
\,\frac{d\epsilon(F)}{dF}\right)\,\delta\,A_{\underline{\lambda\mu\nu}}\, \\[2mm]
&=-
\int_{\mathbb{R}^4} \,d^4x\, \sqrt{|g|}\,\,\frac{\partial}{\partial A_{\underline{\lambda\mu\nu}}}\sqrt{\frac{\,-\,\nabla^{\phantom{(a)}}_{[\kappa}\!\!A^{\phantom{(a)}}_{\lambda\mu\nu]}
\,\nabla\,^{[\kappa}\!\!\,A^{\lambda\mu\nu]}}{4!}}\,\left(\frac{R}{16\pi}\,\frac{dG^{-1}(F)}{dF} +
\,\frac{d\epsilon(F)}{dF}\right)\,\delta\,A_{\underline{\lambda\mu\nu}}\, \\[2mm]
&=-
\int_{\mathbb{R}^4} \,d^4x\, \sqrt{|g|}\,\,\frac{1}{2}\frac{1}{F}\,\frac{\partial}{\partial A_{\underline{\lambda\mu\nu}}}\left(\frac{\,-\,{\nabla^{\phantom{(a)}}_{[\kappa}\!\!A^{\phantom{(a)}}_{\lambda\mu\nu]}
\,\nabla\,^{[\kappa}\!\!\,A^{\lambda\mu\nu]}}}{4!}\right)\,\left(\frac{R}{16\pi}\,\frac{dG^{-1}(F)}{dF} +
\,\frac{d\epsilon(F)}{dF}\right)\,\delta\,A_{\underline{\lambda\mu\nu}}\, \\[2mm]
&=-
\int_{\mathbb{R}^4} \,d^4x\, \sqrt{|g|}\,\,\frac{1}{2}\frac{1}{F}\,\left\{ \,-\,24\,\nabla_\kappa \,\delta\,A_{\underline{\lambda\mu\nu}}\, F^{\kappa\underline{\lambda\mu\nu}}\,-\,24\,g_{\kappa\acute\kappa\,}\,g^{\underline{\lambda}\acute\lambda}\,g^{\underline{\mu}\acute\mu}\,g^{\underline{\nu}\acute\nu}\,g^{\underline{\kappa}\hat\kappa}\,\nabla^{\acute\kappa}\,\delta\,A_{\underline{\lambda\mu\nu}}\,\,F_{\hat\kappa\acute\lambda\acute\mu\acute\nu} \right\}\,\times \\[2mm]
& \phantom{aaaaaaaaaaaaaaaaaa}\times\,\frac{1}{4!}\left(\frac{R}{16\pi}\,\frac{dG^{-1}(F)}{dF} +
\,\frac{d\epsilon(F)}{dF}\right)\, \\[4mm]
&=-
\int_{\mathbb{R}^4} \,d^4x\, \sqrt{|g|}\,\,\frac{2}{2}\cdot\frac{\,\nabla_\kappa\,\delta A_{\underline{\lambda\mu\nu}}\,F^{\kappa\underline{\lambda\mu\nu}}}{-F}\,\left(\frac{R}{16\pi}\,\frac{dG^{-1}(F)}{dF} +
\,\frac{d\epsilon(F)}{dF}\right)\,\,. \\
\end{split}\end{equation}
The last part comes due to definition of derivative and variation.
Finally we get for $\delta S$,
\begin{equation}\label{e:barwq}\begin{split}
\delta S&= -\,\int_{\mathbb{R}^4} \,d^4x\, \sqrt{|g|}\,\frac{F^{\kappa\underline{\lambda\mu\nu}}}{-F}\left\{\frac{R}{16\pi}\,\frac{dG^{-1}(F)}{dF} +
\,\frac{d\epsilon(F)}{dF}\,+\right. \\[2mm]
&\left.\,+\,\frac{1}{8}\frac{dK(F)}{dF}\partial^\alpha F^2 \partial_\alpha F^2\,-\,4\,F\,\nabla^\alpha\left(\frac{1}{8} K(F)\,\partial_\alpha F^2\right)\right\}\,\nabla_\kappa \delta A_{\underline{\lambda\mu\nu}} \\[2mm]
&= -\,\int_{\mathbb{R}^4} \,d^4x\, \sqrt{|g|}\,\nabla_\kappa\,\left\{\frac{F^{\kappa\underline{\lambda\mu\nu}}}{-F}\,\left(\frac{R}{16\pi}\,\frac{dG^{-1}(F)}{dF} +
\,\frac{d\epsilon(F)}{dF}\,+\right. \right.\\[2mm]
&\left.\left.\,+\,\frac{1}{8}\frac{dK(F)}{dF}\partial^\alpha F^2 \partial_\alpha F^2\,-\,4\,F\,\nabla^\alpha\left( \frac{1}{8}K(F)\,\partial_\alpha F^2\right)\right)\, \delta A_{\underline{\lambda\mu\nu}}\right\} \,+\, \\[2mm]
&\,+\,\int_{\mathbb{R}^4} \,d^4x\, \sqrt{|g|}\,\nabla_\kappa\,\left\{\frac{F^{\kappa\underline{\lambda\mu\nu}}}{-F}\,\left(\frac{R}{16\pi}\,\frac{dG^{-1}(F)}{dF} +
\,\frac{d\epsilon(F)}{dF}\,+\right. \right.\\[2mm]
&\left.\left.\,+\,\frac{1}{8}\frac{dK(F)}{dF}\partial^\alpha F^2 \partial_\alpha F^2\,-\,4\,F\,\nabla^\alpha\left( \frac{1}{8}K(F)\,\partial_\alpha F^2\right)\right)\right\}\, \delta A_{\underline{\lambda\mu\nu}}. 
\end{split}\end{equation}
The first part of the sum is zero due gauss and we can set $\delta A$ to zero at far boundary. So we have then
\begin{equation}\label{e:barwq}\begin{split}
\delta S&= \,+\,\int_{\mathbb{R}^4} \,d^4x\, \sqrt{|g|}\,\nabla_\kappa\,\left\{\frac{F^{\kappa\underline{\lambda\mu\nu}}}{-F}\,\left(\frac{R}{16\pi}\,\frac{dG^{-1}(F)}{dF} +
\,\frac{d\epsilon(F)}{dF}\,+\right. \right.\\[2mm]
&\left.\left.\,+\,\frac{1}{8}\frac{dK(F)}{dF}\partial^\alpha F^2 \partial_\alpha F^2\,-\,4\,F\,\nabla^\alpha\left( \frac{1}{8}K(F)\,\partial_\alpha F^2\right)\right)\right\}\, \delta A_{\underline{\lambda\mu\nu}}\,. 
\end{split}\end{equation}
As $\frac{F^{\kappa\lambda\mu\nu}}{F }\,=\,\frac{e^{\kappa\lambda\mu\nu}}{\sqrt{|g|}}\,$ ie. constant, g commutes with $\nabla_\kappa$, and $\delta S\,=\,0$, from above follows
\begin{equation}\label{e:maxwell}\begin{split}
\nabla_\kappa\,\left(\frac{R}{16\pi}\,\frac{dG^{-1}(F)}{dF} +
\,\frac{d\epsilon(F)}{dF}\,+\,\frac{1}{8}\frac{dK(F)}{dF}\partial^\alpha F^2 \partial_\alpha F^2\,-\,\frac{1}{2}\,F\,\nabla^\alpha\left( K(F)\,\partial_\alpha F^2\right)\right)\,=\,0 . 
\end{split}\end{equation}
\\

To get Einstein equations we first variate $S_1$ given in equation \eqref{eq:ActionOpen1} with respect to $\delta g^{\mu\nu}$.
\label{eq:actionmatter1}\\[2mm]
\begin{equation}\label{e:barwq}\begin{split}
S_1\,&+\,\delta S_1 = -\int_{\mathbb{R}^4} \,d^4x\,\left\{\left(\sqrt{-g}\,+\delta\,\sqrt{-g}\right)\frac{1}{8}\,K(F+\delta F)\,
\left(g^{\alpha\beta}\,+\,\delta g^{\alpha\beta}\right)\right. \times \\[2mm]
&\times\,\left. \left(\nabla_{\beta}\,(\,\nabla^{\phantom{(a)}}_{[\kappa}\!\!A^{\phantom{(a)}}_{\lambda\hat\mu\hat\nu]}\,
\,\nabla\,^{[\kappa}\!\!\,A^{\lambda\hat\mu\hat\nu]}\,)\,+\,\delta\left[\nabla_{\beta}\,(\,\nabla^{\phantom{(a)}}_{[\kappa}\!\!A^{\phantom{(a)}}_{\lambda\hat\mu\hat\nu]}\,
\,\nabla\,^{[\kappa}\!\!\,A^{\lambda\hat\mu\hat\nu]}\,)\,\right]\right)\right. \times  \\[2mm]
 &\times\, \left. \left. \left(\nabla_{\alpha}\,(\,\nabla^{\phantom{(a)}}_{[\acute\kappa}\!\!A^{\phantom{(a)}}_{\acute\lambda\acute\mu\acute\nu]}\,
\,\nabla\,^{[\acute\kappa}\!\!\,A^{\acute\lambda\acute\mu\acute\nu]}\,)\,+\,\delta\left[\nabla_{\alpha}\,(\,\nabla^{\phantom{(a)}}_{[\acute\kappa}\!\!A^{\phantom{(a)}}_{\acute\lambda\acute\mu\acute\nu]}\,
\,\nabla\,^{[\acute\kappa}\!\!\,A^{\acute\lambda\acute\mu\acute\nu]}\,)\,\right]\right)\right.\right\}\,\frac{1}{4!^2}\,\delta g^{\mu\nu}\,. \\[2mm]
\end{split}\end{equation} \\[2mm]
First derive $\delta\,\sqrt{-g}$, we have 
\beqa\label{eq:ActionOpen}
\frac{\partial \sqrt{-g}}{\partial g^{\mu\nu}}\,=\,-\,\frac{\partial g}{\partial g^{\mu\nu}}\,\frac{1}{2}\,\frac{1}{\sqrt{-g}}\,=\,\frac{\partial g}{\partial g^{\mu\nu}}\,\frac{1}{2}\,g^{-1}\,\sqrt{-g}\,.
\eeqa
Jacobi relation and relation for differential inversion of a matrix combined gives 
\beqa\label{eq:ActionOpen}
\frac{\partial g}{\partial g^{\mu\nu}}\,=\,-\,\frac{g_{\mu\nu}}{g^{\acute\mu\acute\nu}}\,\frac{\partial g}{\partial g_{\acute\mu\acute\nu}},
\eeqa
and finally 
\beqa\label{eq:ActionOpen}
\delta \sqrt{-g}\,=\,-\,\frac{1}{2}\sqrt{-g}\,g_{\mu\nu}\,{\delta g^{\mu\nu}}.
\eeqa
For  $K(F+\delta F)$ we get
\beqa\label{eq:ActionOpen11}
K(F+\delta F)\,=\,K(F)+\frac{dK(F)}{dF}\,\delta F.
\eeqa
To find $\delta F$ we use definition \eqref{eq:Fdefinition2}
\beqa\label{eq:ActionOpen111}
\delta F_{\kappa\lambda\mu\nu}\,=\,\delta F\,\sqrt{-g}\,e_{\kappa\lambda\mu\nu}\,+\,F\,\delta \sqrt{-g}\,e_{\kappa\lambda\mu\nu},
\eeqa
where first $\delta$ is when metric is constant and we should vary $A$ and second when $\delta A\,=\,0$, thus in this $\delta g^{\mu\nu}$ case the first term is zero. So we get
\beqa\label{eq:ActionOpen1111}
\delta F_{\kappa\lambda\mu\nu}\,=\,-\,\frac{1}{2}\sqrt{-g}\,g_{\alpha\beta}\,\delta g^{\alpha\beta}\,F\,e_{\kappa\lambda\mu\nu}\,=\,-\,\frac{1}{2}\,g_{\alpha\beta}\,\delta g^{\alpha\beta}\,F_{\kappa\lambda\mu\nu},
\eeqa
and similarly
\beqa\label{eq:ActionOpen11111}
\delta F\,=\,-\,\frac{1}{2}\,g_{\alpha\beta}\,\delta g^{\alpha\beta}\,F.
\eeqa
Now we get for $\delta K(F)$
\beqa\label{eq:ActionOpen21111}
K(F+\delta F)\,=\,K(F)\,-\,\frac{1}{2}\,\frac{dK(F)}{dF}\,F\,g_{\alpha\beta}\,\delta g^{\alpha\beta}\,.
\eeqa
Later we need also $\delta F^2$
\begin{equation}\label{e:barwq1111}\begin{split}
-\,4!\,\delta F^2\,=&\,\delta F_{\kappa\lambda\mu\nu}\,g^{\kappa\acute\kappa}g^{\lambda\acute\lambda}g^{\mu\acute\mu}g^{\nu\acute\nu}\,F_{\acute{\kappa}\acute{\lambda}\acute{\mu}\acute{\nu}}\,+\,F_{\kappa\lambda\mu\nu}\,g^{\kappa\acute\kappa}g^{\lambda\acute\lambda}g^{\mu\acute\mu}g^{\nu\acute\nu}\,\delta F_{\acute{\kappa}\acute{\lambda}\acute{\mu}\acute{\nu}}\,+ \\[2mm]
&+\,F_{\kappa\lambda\mu\nu}\delta g^{\kappa\acute\kappa}g^{\lambda\acute\lambda}g^{\mu\acute\mu}g^{\nu\acute\nu}F_{\acute{\kappa}\acute{\lambda}\acute{\mu}\acute{\nu}}\,+\,F_{\kappa\lambda\mu\nu}g^{\kappa\acute\kappa}\delta g^{\lambda\acute\lambda}g^{\mu\acute\mu}g^{\nu\acute\nu}F_{\acute{\kappa}\acute{\lambda}\acute{\mu}\acute{\nu}}\,+\\[2mm]
&+\,F_{\kappa\lambda\mu\nu}g^{\kappa\acute\kappa}g^{\lambda\acute\lambda}\delta g^{\mu\acute\mu}g^{\nu\acute\nu}F_{\acute{\kappa}\acute{\lambda}\acute{\mu}\acute{\nu}}\,+\,F_{\kappa\lambda\mu\nu}g^{\kappa\acute\kappa}g^{\lambda\acute\lambda}g^{\mu\acute\mu}\delta g^{\nu\acute\nu}F_{\acute{\kappa}\acute{\lambda}\acute{\mu}\acute{\nu}}\,=\\[2mm]
=&\,-\,\frac{1}{2}\,g_{\alpha\beta}\,\delta g^{\alpha\beta}\,F^2\,-\,\frac{1}{2}\,g_{\alpha\beta}\,\delta g^{\alpha\beta}\,F^2\,+\,F_{\kappa\lambda\mu\nu}\delta g^{\kappa\acute\kappa}F_{\acute{\kappa}}^{\phantom{a}\lambda\mu\nu}\,+\,...\,=\\[2mm]
=&\,-\,g_{\alpha\beta}\,\delta g^{\alpha\beta}\,F^2\,+\,\frac{1}{4}\,g_{\acute{\kappa}\kappa}\,\delta g^{\kappa\acute{\kappa}}\,F^2\,+\\[2mm]
&+\,\frac{1}{4}\,g_{\acute{\lambda}\lambda}\,\delta g^{\lambda\acute{\lambda}}\,F^2\,+\,\frac{1}{4}\,g_{\acute{\mu}\mu}\,\delta g^{\mu\acute{\mu}}\,F^2\,+\,\frac{1}{4}\,g_{\acute{\nu}\nu}\,\delta g^{\nu\acute{\nu}}\,F^2\,=\,0\,.
\end{split}\end{equation}
For $\delta\,(\nabla_\alpha F^2)$ we get
\begin{equation}\label{e:bas1111}\begin{split}
\delta (\nabla_\alpha F^2)\,=&\delta \left(\nabla_{\alpha}\,\left(\frac{\,\nabla^{\phantom{(a)}}_{[\kappa}\!\!A^{\phantom{(a)}}_{\lambda\mu\nu]}\,
\,\nabla\,^{[\kappa}\!\!\,A^{\lambda\mu\nu]}}{-4!}\,\right)\,\right)\,
\,=\,\nabla_\alpha \delta F^2\,+\,\delta\,\nabla_{\alpha}\,\left(\frac{\,\nabla^{\phantom{(a)}}_{[\kappa}\!\!A^{\phantom{(a)}}_{\lambda\mu\nu]}\,
\,\nabla\,^{[\kappa}\!\!\,A^{\lambda\mu\nu]}}{-4!}\,\right)\,=\\[2mm]
=&\,\delta\,\nabla_{\alpha}\,(\,\nabla^{\phantom{(a)}}_{[\kappa}\!\!A^{\phantom{(a)}}_{\lambda\mu\nu]}\,)\,
\frac{\nabla\,^{[\kappa}\!\!\,A^{\lambda\mu\nu]}}{-4!}+\,\frac{\nabla^{\phantom{(a)}}_{[\kappa}\!\!A^{\phantom{(a)}}_{\lambda\mu\nu]}}{-4!}\,
\delta\,\nabla_{\alpha}\,\left(\,\nabla\,^{[\kappa}\!\!\,A^{\lambda\mu\nu]}\,\right)\,.
\end{split}\end{equation}
as $\delta \nabla$ commutes with metric part. Using formula for $\delta\,\nabla_a\nabla_b$ we get 
\begin{equation}\label{e:bas1111}\begin{split}
\delta\,\nabla_{\alpha}\,\nabla^{\phantom{(a)}}_{[\kappa}\!\!A^{\phantom{(a)}}_{\lambda\mu\nu]}\,
\,=\,-\,\frac{1}{2}\,F_{\kappa\lambda\mu\nu}\,\nabla^c\,\delta g_{\alpha c}\,,
\end{split}\end{equation}
and further 
\begin{equation}\label{e:bas1211}\begin{split}
\delta\,\nabla_{\alpha}\,\nabla^{\phantom{(a)}}_{[\kappa}\!\!A^{\phantom{(a)}}_{\lambda\mu\nu]}\,\nabla\,^{[\kappa}\!\!\,A^{\lambda\mu\nu]}\,\frac{1}{-4!}
\,=\,+\,\frac{1}{2}\,F^2\,\nabla^c\,\delta g_{\alpha c}\,,
\end{split}\end{equation}
Summing up all terms (2x4/4) we get finally for $\delta\,(\nabla\,F^2)$
\begin{equation}\label{e:bas1211}\begin{split}
\delta\,(\nabla\,F^2)\,=\,-\,F^2\,g_{\mu\nu}\,\delta\,g^{\mu\nu}\,\nabla_\alpha\,.
\end{split}\end{equation}
Next we gather all results together and get 
\begin{equation}\label{e:barwq11}\begin{split}
\delta S_1 =& -\int_{\mathbb{R}^4} \,d^4x\,\sqrt{-g}\,\,\frac{1}{8}\left\{\,-\frac{1}{2}\,K(F)\,\nabla_\alpha F^2\,\nabla^\alpha F^2\,g_{\mu\nu} + \right. \\[2mm]
&\left.+\,\frac{dK(F)}{dF}\,(-\,\frac{1}{2})\,F\,\,\nabla_\alpha F^2\,\nabla^\alpha F^2\,g_{\mu\nu}\,+\,K(F)\,\,\nabla_\mu F^2\,\nabla_\nu F^2\,+ \right. \\[2mm]
&\left.-\,2\,K(F)\,F^2\,\nabla_\alpha \nabla^\alpha F^2\,g_{\mu\nu} \right\}\,\delta g^{\mu\nu}\,. \\[2mm]
\end{split}\end{equation} \\[2mm]

$\delta S_0$ can be calculated as in \cite{KV2008b} before equation (2.3). We now combine all results (and remembering that  $\delta S_0$ has been multiplied by a factor 2), and demand that $\delta S\,=\,0$ for any $\delta g^{\mu\nu}$. The generalized Einstein equations become
\begin{equation}\label{e:einstein1112}\begin{split}
& \frac{1}{8\pi G(F)}\Big( R_{\mu\nu}-\half\,R\,g_{\mu\nu} \Big)\,+\,\frac{1}{16\pi}\,F\,\frac{dG^{-1}}{dF}\,R\,g_{\mu\nu}\,+\,\frac{1}{8\pi}
\Big( \nabla_\mu\nabla_\nu\, G^{-1}(F) - g_{\mu\nu}\, \Box\, G^{-1}(F)\Big)\,+ \\[2mm]
& -\left(\epsilon (F)-F\,\frac{d\epsilon (F)}{dF}\,\right)g_{\mu\nu}\,\,-\,\frac{1}{8}\,K(F)\,\nabla_\alpha F^2\,\nabla^\alpha F^2\,g_{\mu\nu}\,-\,\frac{1}{8}\,\frac{dK(F)}{dF}\,F\,\,\nabla_\alpha F^2\,\nabla^\alpha F^2\,g_{\mu\nu}\,+\\[2mm]
&-\,\frac{1}{8}\cdot 2\,K(F)\,\,\nabla_\mu F^2\,\nabla_\nu F^2\,-\,\frac{1}{8}\cdot 4\,K(F)\,F^2\,\nabla_\alpha \nabla^\alpha F^2\,g_{\mu\nu}\,=0\,.
\end{split}\end{equation}
From Maxwell equation  \eqref{e:maxwell} we get 
\begin{equation}\label{e:maxwell1112}\begin{split}
\frac{R}{16\pi}\,\frac{dG^{-1}(F)}{dF} +
\,\frac{d\epsilon(F)}{dF}\,+\,\frac{1}{8}\frac{dK(F)}{dF}\partial^\alpha F^2 \partial_\alpha F^2\,-\,\frac{1}{2}\,F\,\nabla^\alpha\left( K(F)\,\partial_\alpha F^2\right)\,=\mu, 
\end{split}\end{equation}
where $\mu$ is some constant. Now we set 
\begin{equation}\label{e:111}\begin{split}
C(F)\,=\,F^2\,K(F).
\end{split}\end{equation}
This gives then 
\begin{equation}\label{e:maxwell11112}\begin{split}
\frac{R}{16\pi}\,\frac{dG^{-1}(F)}{dF} +
\,\frac{d\epsilon(F)}{dF}\,-\, \frac{1}{2}\frac{dC(F)}{dF}\,\partial^{\alpha} F \partial_{\alpha} F\,-\, C(F) \,\Box F\,\,=\mu. 
\end{split}\end{equation}
For $\delta S_1$ we get when we substitute $C(F)$ and use again $\nabla_\alpha F$ instead of $\nabla_\alpha F^2$
\begin{equation}\label{e:barwq2211}\begin{split}
\delta S_1 =& -\int_{\mathbb{R}^4} \,d^4x\,\sqrt{-g}\,\,\frac{1}{8}\left\{\,-\frac{1}{2}\,C(F)\cdot 4\cdot\nabla_\alpha F\,\nabla^\alpha F\,g_{\mu\nu} + \right. \\[2mm]
&\left.-\,\frac{1}{2}\,\frac{dC(F)}{dF}\,\,F\cdot 4\cdot\nabla_\alpha F\,\nabla^\alpha F\,g_{\mu\nu}\,+\,\frac{1}{2}\,F\,\frac{2}{F}C(F)\cdot 4\cdot\nabla_\alpha F\,\nabla^\alpha F\,g_{\mu\nu}\,+ \right. \\[2mm]
& \left.+\,C(F)\cdot 4\cdot\nabla_\mu F\,\nabla_\nu F\,-\,\frac{2\cdot 2\,F^2}{F^2}C(F)\,\nabla_\alpha F\,\nabla^\alpha F\,g_{\mu\nu}\,+ \right. \\[2mm]
&\left.\,-\,\frac{2\cdot 2\,F^2}{F^2}\,F\,C(F)\,\nabla_\alpha \nabla^\alpha F\,g_{\mu\nu} \right\}\,\delta g^{\mu\nu}\,. \\[2mm]
\end{split}\end{equation} \\[2mm]
For Einstein equation \eqref{e:einstein1112} we get now
\begin{equation}\label{e:einstein111}\begin{split}
& \frac{1}{8\pi G(F)}\Big( R_{\mu\nu}-\half\,R\,g_{\mu\nu} \Big)\,+\,\frac{1}{16\pi}\,F\,\frac{dG^{-1}}{dF}\,R\,g_{\mu\nu}\,+ \\[2mm]
& +\,\frac{1}{8\pi}
\Big( \nabla_\mu\nabla_\nu\, G^{-1}(F) - g_{\mu\nu}\, \Box\, G^{-1}(F)\Big)\,-\left(\epsilon (F)-F\,\frac{d\epsilon (F)}{dF}\,\right)g_{\mu\nu}\,+ \\[2mm]
&-\,\frac{1}{2}\,C(F)\,g_{\mu\nu}\,
\nabla_{\alpha} F \,\nabla^{\alpha} F \,-\,\frac{1}{2}\,F\,\frac{dC(F)}{dF}\,\nabla_{\alpha} F \,\nabla^{\alpha} F\,g_{\mu\nu}\,+ \\[2mm]
&+ \,\,C(F)
\nabla_{\mu} F \,\nabla_{\nu} F\,-\,C(F)\,\Box F\,g_{\mu\nu}\,=0\,.
\end{split}\end{equation}
From equation \eqref{e:maxwell1112} we can write
\begin{equation}\label{e:maxwell112}\begin{split}
\frac{d\epsilon(F)}{dF}\,=\,-\,\frac{R}{16\pi}\,\frac{dG^{-1}(F)}{dF} + \frac{1}{2}\frac{dC(F)}{dF}\,\nabla_{\alpha} F \nabla^{\alpha} F\,+\, C(F) \,\Box F\,+\,\mu. 
\end{split}\end{equation}
Substituting the above equality we finally get a simplified form of Einstein equations
\begin{equation}
\label{e:einstein1111}
\begin{split}
& \frac{1}{8\pi G(F)}\Big( R_{\mu\nu}-\half\,R\,g_{\mu\nu} \Big)\,+\,\frac{1}{8\pi}
\Big( \nabla_\mu\nabla_\nu\, G^{-1}(F) - g_{\mu\nu}\, \Box\, G^{-1}(F)\Big)\,+ \\[2mm]
& -\left(\epsilon (F)-\,\mu\,F\,\right)g_{\mu\nu}\,-\,\frac{1}{2}\,C(F)\,g_{\mu\nu}\,
\nabla_{\alpha} F \,\nabla^{\alpha} F \,+\,C(F)
\nabla_{\mu} F \,\nabla_{\nu} F\,=0\,.
\end{split}
\end{equation}

%%%%%%%%%%%%

\end{document}